\documentclass[12pt,fleqn]{article}

\usepackage{graphicx}
\usepackage{dcolumn}
\usepackage{bm}
\usepackage{amsmath}
\usepackage{amssymb}  
\usepackage{subfigure}
\usepackage{psfig,float}
\def\OMIT#1{}
\begin{document} 
 
\title{Symmetry and Dynamics in living organisms: The self-similarity principle governs gene expression dynamics }

\author{T. Ochiai{\footnote{These authors contributed equally to this work. \newline
Corresponding authors: ochiai@kuicr.kyoto-u.ac.jp, nacher@kuicr.kyoto-u.ac.jp
}}, J.C. Nacher{\footnotemark[1]}, T. Akutsu}     
 
\maketitle
 
\begin{center}
{\it Bioinformatics Center, Institute for Chemical Research, Kyoto University, }\end{center}
\begin{center}
{\it Uji, 611-0011, Japan}
\end{center}

\begin{center}
\end{center} 

\begin{center}

\end{center}

\begin{abstract}
{\small{
The ambitious and ultimate research purpose in Systems Biology is the understanding 
and modelling of the cell's system. Although a vast number of models have been 
developed in order to extract biological knowledge from complex systems composed 
of basic elements as proteins, genes and chemical compounds, a need remains for 
improving our understanding of dynamical features of the 
systems (i.e., temporal-dependence). 


In this article, we analyze the gene expression dynamics (i.e., how 
the genes expression fluctuates in time) by using a new constructive 
approach. This approach is based on only two fundamental ingredients: symmetry and 
the Markov property of dynamics. First, by using experimental data of human and 
yeast gene expression time series, we found a symmetry in short-time transition 
probability from time $t$ to time $t+1$. We call it self-similarity 
symmetry (i.e., surprisingly, the gene expression short-time 
fluctuations contain a repeating pattern of smaller and 
smaller parts that are like the whole, but different in size). Secondly, the 
Markov property of dynamics reflects that the short-time fluctuation governs 
the full-time behaviour of the system. Here, we succeed in reconstructing 
naturally  the global behavior of the observed distribution of gene 
expression (i.e., scaling-law) and the local behaviour of the 
power-law tail of this distribution, by using only these two 
ingredients: symmetry and the Markov property of dynamics. This approach 
may represent a step forward toward an integrated image of the basic 
elements of the whole cell. 


}}

\end{abstract}


\section{Introduction}

The final goal in Systems Biology is the understanding 
and modelling of the cell's system. In the cell, the expression level of genes plays a key role, since gene expression 
is a complex transcriptional process where {\it mRNA} molecules are translated into proteins, which control 
most of the cell functions. Recently, gene expression profiles for different types of cells of
several organisms have been measured, and some experiments have also provided data about the fluctuation of expression level of thousands genes
in time. Here, we propose 
a stochastic approach to gain insight into gene expression time series data, in order to uncover fundamental principles that govern the
gene dynamics.

Although many complex systems may be governed 
by non-stochastic processes, in the gene expression problem the random variation is reasonable,
plays a relevant role in cellular process, and furthermore stochastic noise (e.g., {\it intrinsic and extrinsic}) 
have recently been measured and studied theoretically \cite{elo,
paulsson, blake, hasty, sato}. For example, 
the expression level of thousands genes is very 
low, which
creates intrinsic uncertainties in the number of expressed genes in the cells \cite{kuz}. 
Furthermore, the number 
of molecules which are involved in signal transduction pathways fluctuates
from $10^2$ to $10^4$. Therefore, the randomness connected with elementary molecular interactions and their amplification 
in the signaling cascade generates significant spatio-temporal noise. Therefore, the stochastic approach seems more
plausible than the deterministic approach. Finally, we also note that the current experimental
 techniques also generate an additional source of fluctuation, which come from the ubiquitous instrumental noise
  (which may be around $30\%$ or more) from chip to chip with the current GeneChips technologies.

Currently, DNA microarray/GeneChips \cite{chip1,chip2} (like the pendulum 
clock of Hyugens used by Newton to uncover the dynamical laws written 
in his famous {\it Principia Mathematica}) offers 
the ability to monitor in time changes in expression level of large 
subsets of genes from a variety of organisms on a scale unattainable by 
other methods. In particular, experiments 
 done on time series of absolute 
value of gene expression level studying the yeast mitotic cell cycle \cite{yeast}  
and transcriptional regulation during human cell cycle \cite{human} have provided a 
huge wealth of data to uncover general principles of gene expression dynamics.

\begin{figure}[htb]
\setlength{\unitlength}{1cm}
\begin{picture}(15,12)(-1,-1)
\put(-1,0){\includegraphics[scale=0.5]{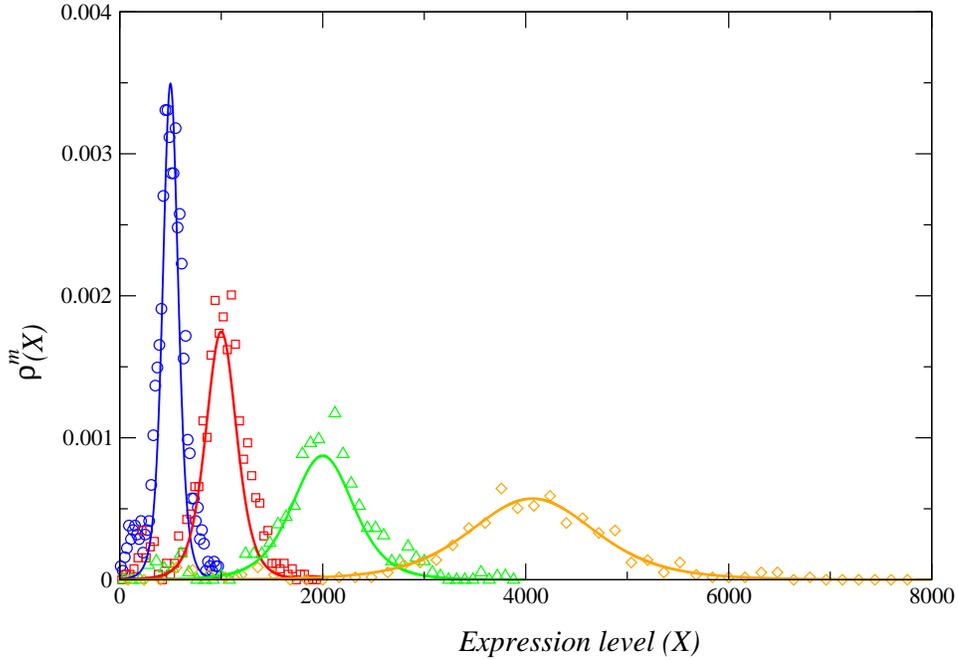}}
\end{picture}  
\caption{\small{We show the distribution $\rho^m(x)$. It shows the distribution of expression level $x$ of genes with average expression level $m$. Horizontal axis denotes the gene expression level $x$. Vertical axis denotes the distribution $\rho^m(x)$ of genes which fluctuate around mean value $m$. 
 Continuous line shows the prediction of our model by using the probability distribution  (Eq. (\ref{eqn: solution of gene SPDE})). Dots show experimental data of Human organism from \cite{human}. Orange: genes with fluctuations around expression value of $m$=4000 (resp., Green: 2000. Red: 1000. Blue: 500. ) This figure shows the scaling law: $\rho^{cm}(x)=\frac{1}{c}\rho^m(x/c)$ (More symmetric form: $\rho^{cm}(x)dx=\rho^m(x/c)d(x/c)$) for arbitrary real number $c$. 
 This figure is constructed from selected figures with $m=500, 1000, 2000$ and $4000$ in Fig \ref{fig: distribution human 1}, \ref{fig: distribution human 2}.
}}
\label{fig: probability distribution}
\end{figure}

 In this article, by using these experimental data \cite{yeast,human}, we have 
 found the following interesting phenomena: 1) "Mean- Reverting" and "Extreme Value- Jumps-More" 
 mechanisms. These mechanisms are present in short-time 
 fluctuation (i.e., how the gene expression level changes 
 from time-step $t$ to time-step $t+1$). Surprisingly, these two mechanisms 
 are governed by a symmetry: Self-similarity (see Fig. \ref{fig: Mean-Reverting} and Fig. \ref{fig: Extreme-Value-Jumps-More}).  By inserting this 
 information (i.e., two mechanisms and the self-similarity symmetry) into the
  general stochastic partial differential equation (SPDE) which 
  spontaneously emerges from Markov property, we found the fundamental 
  equation for gene expression dynamics. This equation has a high predictive power. Here, we 
  enumerate a couple of results. For example, the distribution 
  solution of this equation  $\rho^m(x)$ gives the distribution of expression level $x$ of genes which fluctuate around $m$ (i.e., their average expression is in the vicinity of 
  the value $m$).  This solution agrees with the 
  experimental data shown later in Fig. \ref{fig: probability distribution} (for more  
  details see Figs. \ref{fig: distribution human 1}, \ref{fig: distribution human 2} and \ref{fig: distribution yeast}). Furthermore, we rebuild   
  the observed global behaviour of gene expression distribution, which is 
  characterized by the following relation $\rho^{cm}(x)=\frac{1}{c}\rho^{m}(x/c)$  (or as the more symmetric form $\rho^{cm}(x)dx=\rho^{m}(x/c)d(x/c)$). We call it scaling-law formula.  This formula 
  indicates that distribution of gene expression level also has the self-similarity 
  structure (We can observe a repeating pattern of smaller and 
smaller parts that are like the whole, but different in size in Figs. \ref{fig: distribution human 1}, \ref{fig: distribution human 2} and \ref{fig: distribution yeast}), which is a direct 
  consequence of the self-similarity symmetry in short-time 
  transition probability (Fig. \ref{fig: Mean-Reverting} and Fig. \ref{fig: Extreme-Value-Jumps-More}). In a sense, the self-similarity 
  governs all the dynamical structure of gene expression 
  phenomena, by means of Markov property. On the 
  other hand, our model can predict sensitive 
  aspects of gene expression systems. More precisely, our constructive 
  model predicts that the tail of the distribution has a power-law tail $\rho^m(x)\propto x^{-4}$ (when $x\to \infty$), which is observed in experimental data (see Fig. \ref{fig: Normal vs power}).

\section{Databases} 
\paragraph{Databases.}
We used experimental data from two well-known experiments on Yeast \cite{yeast} and 
Human \cite{human}  organisms. The first experiment analysed cell cycle of the budding yeast {\it S. Cerevisiae}, where 
around 6220 genes were monitored. Data of absolute value of gene expression fluctuations were 
collected at 17 time points every 10 min intervals. Data was obtained from WWW site http://genomics.standford.edu.

The second experiment identifies cell-cycle-regulated transcripts in human cells 
using high-density oligonucleotide arrays. Data are collected every 2 hours for 24 hours, what is 
equivalent to almost 2 full cell cycles. The number of genes monitored was around 35000, and data for 
absolute value of gene expression fluctuations was obtained \\
from WWW site http://www.salk.edu/docs/labs/chipdata.

\begin{figure}[htb]
\setlength{\unitlength}{1cm}
\begin{picture}(15,12)(-1,-1)
\put(-1,0){\includegraphics[scale=0.65]{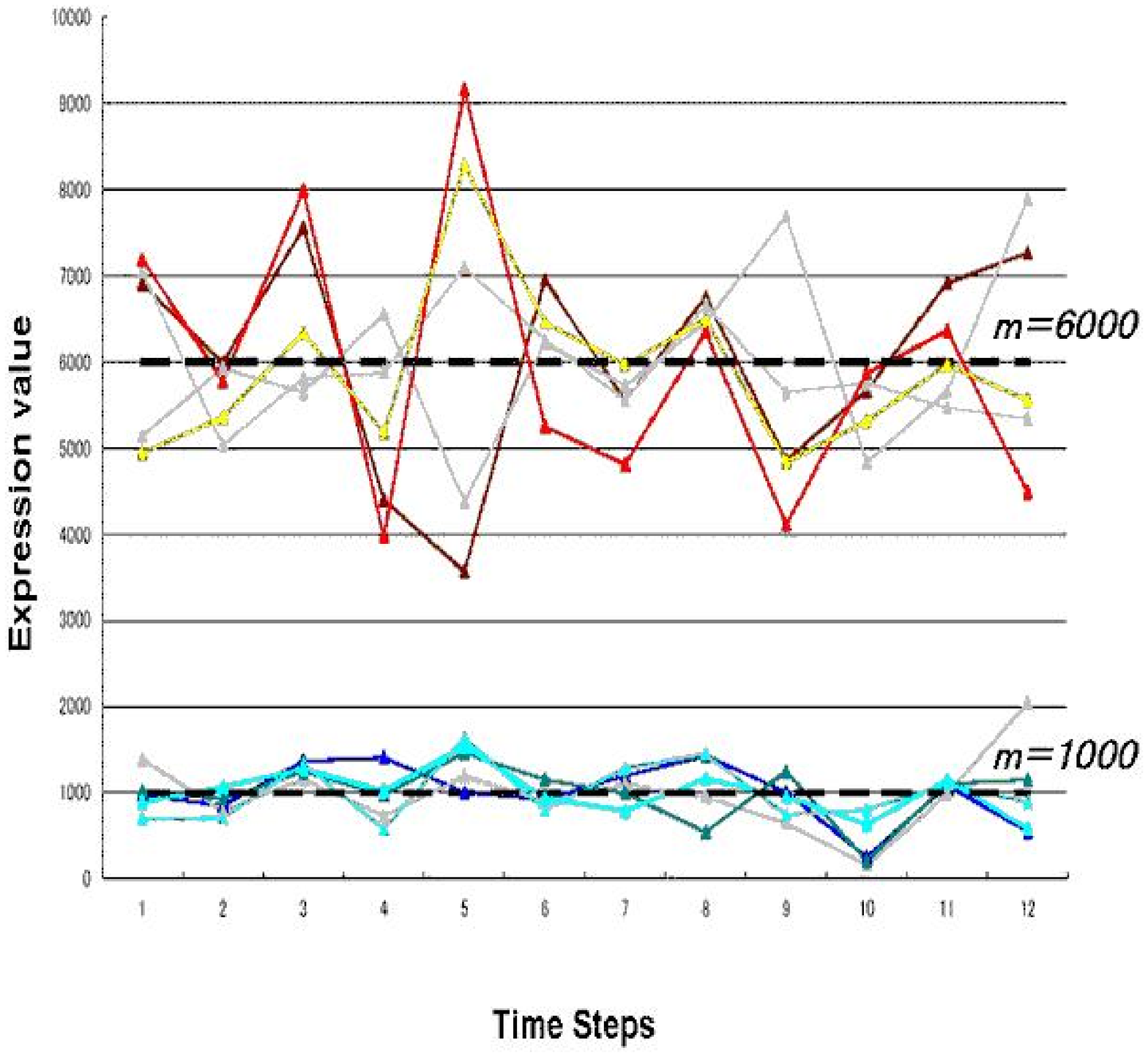}}
\end{picture} 
\caption{\small{As an example, we show the gene expression time series data of a few genes which belong to human organism \cite{human}. We see that the gene expression value fluctuates around the mean value $m=6000$ and $m=1000$. Therefore, we group all these genes into two subgroups (one group consists of all genes fluctuating around $m=6000$ and the other group consists of all genes fluctuating around $m=1000$) according to the mean value $m$.}}
\label{fig: time series}
\end{figure}

\paragraph{Subgroups of Genes.}  In order to explain in more detail our findings, let us observe Fig. \ref{fig: time series}. We see that gene  
expression time series for genes fluctuates around some mean value. Therefore, it seems 
natural to classify all the genes of human and yeast organisms into subgroups according 
to the time-averaged expression value of each gene denoted by $m$.  We cluster the data of 
genes by using the mean value $m$. We made a group of genes composed by all the genes 
within 1\% of the following values of $m$: 500, 1000, 1500, 2000. Furthermore, all the genes which are 
within 1\% of the previous values of $m$ are called {\it subgroups of genes}. We also make more groups 
and subgroups of genes. We detail them as follows: within 2\% of the values of ($m$=2500, 3000), within 3\% 
of the values of ($m$=3500, 4000, 4500), within 4\% of the values of ($m$=5000, 5500, 6000) and within 5\%  
of the values of ($m$=7000, 8000, 9000).

\section{Methods}
Let $\{X(t), 0 \le t <\infty \}$ be a stochastic process. For $(s>t)$ , the conditional probability 
density function $p(y,s|x,t)=p(X(s)=y | X(t)=x )$ is defined as usual manner. For the matter of 
convenience, we often write  $p(y,s)$ for $p(y,s|x,t)$ .
\paragraph{Ito Stochastic Process.} By observing Fig. \ref{fig: time series}, we see that the gene expression time series fluctuate around their 
own mean value, which is denoted by $m$. Therefore, we characterized the gene 
expression time series data by the mean value $m$. For each set of genes which fluctuates around the mean value $m$, we use the most general Stochastic 
Partial Differential Equation (SPDE) (see \cite{ochi,ochi2,Kampen,Wong,black} for more details). More precisely, we respectively use the following general SPDE for each gene which fluctuates around mean value $m$ (gene whose average expression level is $m$):
\begin{eqnarray}\label{eqn: general SPDE}
dX^m(t)=\alpha^m(X^m(t)) dt + \beta^m(X^m(t))dW(t),
\end{eqnarray}
where stochastic variable $X^m(t)$ denotes the gene expression level which has mean value $m$ and $W(t)$ denotes the Wiener process . Here $\alpha^m(x)$ denotes the 
drift (i.e., by starting with expression level $x$ at time $t$, it means the average change of expression level $x$ at time $t+\epsilon$) defined by     
\begin{eqnarray}\label{eqn: initial condition alpha}
\alpha^m(x)=\lim_{\epsilon \to 0}\frac{1}{\epsilon}\int_{-\infty}^\infty (y-x) T_\epsilon^m(y,x) dy,
\end{eqnarray}                      
$\beta^m(x)$ denotes the diffusion   (i.e., by starting with the expression level $x$ at time $t$, it means the average 
jumping size of expression level $x$ at time $t+\epsilon$) defined by
\begin{eqnarray}\label{eqn: initial condition beta}
\beta^m(x)=\Bigl(\lim_{\epsilon \to 0}\frac{1}{\epsilon}\int (y-x)^2 T_\epsilon^m(y,x) dy\Bigr)^{\frac{1}{2}}.
\end{eqnarray}                                 
Here, $T_\epsilon^m(y,x)$ is the short-time transition 
probability of genes fluctuating around $m$ defined by $T_\epsilon^m(y,x)=p^m(y,t+\epsilon|x,t)$ for sufficient small $\epsilon$. Details 
of the proof can be found in \cite{ochi,ochi2,Kampen,Wong,black}.
This equation allows us to predict long-time behaviour of gene expression dynamics, 
by using the short-time transition probability  as follows.

\paragraph{How to use the SPDE Eq. (\ref{eqn: general SPDE}).} We can obtain the dynamical information 
of gene expression from experimental data of the short-time transition probability $T_\epsilon^m(y,x)$ ($\epsilon$ is sufficiently small and fixed), by the  
following procedure:
 
\begin{itemize}

\item[{\it (i)}] Given the experimental data of the short-time transition 
probability, we obtain the drift $\alpha^m(x)$ and the diffusion $\beta^m(x)$ by using Eq. (\ref{eqn: initial condition alpha}) and (\ref{eqn: initial condition beta}) respectively.
\item[{\it (ii)}]By inserting the drift $\alpha^m(x)$ and the diffusion $\beta^m(x)$ obtained in previous step into 
the most general SPDE (\ref{eqn: general SPDE}), we obtain the specific equation for gene expression dynamics.
\item[{\it (iii)}]	Solving this specific equation obtained in step two, we can obtain the dynamical information of gene expression.
\end{itemize}

\section{Self-Similarity Symmetry}
Next, for each set of genes which fluctuates around the mean value $m$, we compute the two most important quantities to 
characterize the gene expression fluctuations: 1) the drift  $\alpha^m(x)$ and 2) the diffusion  $\beta^m(x)$ from the experimental data \cite{yeast,human}.

\paragraph{Input data: Short-time transition probability.}   From the 
experimental data \cite{yeast,human}, we compute $T_\epsilon^m(y,x)$ for each set of genes which has mean value $m$. Next, from $T_\epsilon^m(y,x)$, we evaluate the drift $\alpha^m(x)$ and the diffusion $\beta^m(x)$ by using 
Eqs. (\ref{eqn: initial condition alpha}) and (\ref{eqn: initial condition beta}). We show the results of the drift $\alpha^m(x)$ and the diffusion $\beta^m(x)$ in Fig. \ref{fig: Mean-Reverting} and Fig. \ref{fig: Extreme-Value-Jumps-More}
respectively. (More precisely, see Fig. \ref{fig: Mean-Reverting human}, \ref{fig: Mean-Reverting yeast}, \ref{fig: Extreme-Value-Jumps-More human}, \ref{fig: Extreme-Value-Jumps-More yeast}). By observing 
these figures, we see that the drift $\alpha^m(x)$ and the diffsion $\beta^m(x)$ can be fitted by the following analytical expressions:
\begin{eqnarray}\label{eqn: drift}
\alpha^m(x)=\mu(m-x)
\end{eqnarray}   
\begin{eqnarray}\label{eqn: diffusion}
\beta^m(x)=m((x/m-1)^2+b)
\end{eqnarray}            
where $\mu=1$ (resp. $\mu=0.8$) and $b=0.2\sim 0.3$ (resp. $b=0.2\sim 0.3$) for human (resp. yeast) organism.
                    
These results obtained by using experimental data are interesting, 
since expression level of genes follows the same tendency independently of the 
scale $m$ of gene expression (self-similar symmetry in the drift $\alpha^m(x)$  and the diffusion $\beta^m(x)$). In other words, the drift $\alpha^m(x)$  
and the diffusion $\beta^m(x)$ in Fig. \ref{fig: Mean-Reverting} and Fig. \ref{fig: Extreme-Value-Jumps-More} and Figs. \ref{fig: Mean-Reverting human}, \ref{fig: Mean-Reverting yeast}, \ref{fig: Extreme-Value-Jumps-More human}, \ref{fig: Extreme-Value-Jumps-More yeast} contain a repeating pattern of smaller and smaller parts 
that are like the whole, but different in size. In addition, same feature has been found in 
simple organism as Yeast and in complex one as Human. Therefore, this tendency probably 
represents a universal property of gene expression dynamics in all living organisms.

\paragraph{Mean-Reverting mechanism} For the drift $\alpha^m(x)$ (Eq. (\ref{eqn: drift})) computed by Eq. (\ref{eqn: initial condition alpha}), we observe the "Mean-Reverting" mechanism in short-time fluctuation for Human organism in Fig. \ref{fig: Mean-Reverting} (see Figs. \ref{fig: Mean-Reverting human}, \ref{fig: Mean-Reverting yeast} for more details). In other words, genes with higher expression 
 level than the mean value $m$ at time-step $t$ (i.e., $m<x$) tend to decrease their expression level at 
 time-step $t+1$ (i.e., $ \alpha^m(x) < 0$), while genes with lower expression level than the mean 
 value $m$ (i.e., $x<m$) have tendency to increase their expression level (i.e.,  $\alpha^m(x) >0$ ). Interestingly, the parameter $(\mu=1)$ in Eq. (\ref{eqn: drift}) is conserved for all the 
 values $m$ in human organism (i.e., constant for all the subgroups of genes). Surprisingly, a value in the vicinity of one $(\mu=0.8)$  is also observed for   yeast organism (see Fig.\ref{fig: Mean-Reverting yeast}). Futhermore, this value is conserved for all the values $m$ in yeast organism.

\paragraph{Extreme-Value-Jumps-More mechanism} For the diffusion $\beta^m(x)$ (Eq. (\ref{eqn: diffusion})) computed by Eq. (\ref{eqn: initial condition beta}), we observe the "Extreme-Value-Jumps-More" mechanism in short-time 
fluctuation for Human organism in Fig. \ref{fig: Extreme-Value-Jumps-More} (see Figs. \ref{fig: Extreme-Value-Jumps-More human}, \ref{fig: Extreme-Value-Jumps-More yeast} for more details). This mechanism means that values far from the mean 
value $m$ change more. Remarkably, 
the parameter $b=0.2\sim 0.3$ in Eq. (\ref{eqn: diffusion}) is conserved among organisms (same for yeast and human) and among all the 
subgroups of genes of each organism.  (see Figs. \ref{fig: Extreme-Value-Jumps-More human}, \ref{fig: Extreme-Value-Jumps-More yeast})
 
\paragraph{Self-Similarity Symmetry}  The most important fact on short-time transition 
probability analysis is that both $\alpha^m(x)$ and $\beta^m(x)$ show the same pattern for all values 
of  "$m$" (i.e., even although the system is re-scaled by different "$m$" ). We call this 
phenomena self-similarity symmetry of gene expression system. Namely, the gene expression short-time fluctuations contain a repeating pattern of smaller and smaller parts that are 
like the whole, but different in size (See Fig. \ref{fig: Mean-Reverting}, \ref{fig: Extreme-Value-Jumps-More} and Fig. \ref{fig: self-similarity}). In an equation level, we can see this phenomena as 
\begin{eqnarray} 
&&\alpha^{cm}(x)=c \alpha^{m}(x/c) \label{eq: symmetry drift}\\
&&\beta^{cm}(x)=c \beta^{m}(x/c) \label{eq: symmetry diffuion}
\end{eqnarray}
for arbitrary real number $c$.  We remark that the 
self-similarity structure exists in both human and yeast organism. Probably, this symmetry is 
conserved among all the organism. It is also important to remark that this symmetry has been 
revealed by analyzing only experimental data of yeast and human time series.

\section{ The fundamental equation for gene expression dynamics}
\paragraph{The fundamental equation for gene expression dynamics} By substituting these 
expressions  Eqs. (\ref{eqn: drift}) and (\ref{eqn: diffusion}) into Eq. (\ref{eqn: general SPDE}), we obtain:
\begin{eqnarray}\label{eqn: gene SPDE}
dX^m(t)=\mu(m-X^m(t))dt + m((X^m(t)/m-1)^2+b)dW(t),
\end{eqnarray} 
This equation describes how gene expression level changes in time. We believe that it is {\it the fundamental equation for gene expression dynamics}. By looking at this equation, we can see that it contains a relevant 
symmetry. Namely, when we take different $m^\prime$, we can recover the original equation by substituting  $X^{m^\prime}(t)\to X^m(t)m^\prime/m$. This symmetry is a consequence of the self-similarity symmetry of $\alpha^m(x)$ and $\beta^m(x)$ in Eqs. (\ref{eq: symmetry drift}) and (\ref{eq: symmetry diffuion}). 

\paragraph{The solution of the SPDE (\ref{eqn: gene SPDE})} This dynamical equation (\ref{eqn: gene SPDE}) has a powerful prediction 
capability as follows. First, although we can solve it by obtaining a time-dependent 
solution, we focus our attention on the stationary solution. The stationary 
distribution solution of Eq. (\ref{eqn: gene SPDE}) is given by the following equation:
\begin{eqnarray}\label{eqn: solution of gene SPDE}
\rho^m(x)=\frac{K}{m((x/m-1)^2+b)^2}\exp\Bigl(\frac{\mu}{(x/m-1)^2+b} \Bigr)   
\end{eqnarray}
where $K$ is the normalization constant. 
This distribution solution (\ref{eqn: solution of gene SPDE}) gives us the probability distribution that genes fluctuating around $m$ (i.e., its average expression value is $m$) have expression level $x$. Comparing with the experimental 
results, the solution of our model Eq. (\ref{eqn: solution of gene SPDE}) shows a good agreement with experimental data shown in Fig. \ref{fig: probability distribution} (For more  
  details see Figs. \ref{fig: distribution human 1}, \ref{fig: distribution human 2} and \ref{fig: distribution yeast}). 
  
\section{Other Results}
\begin{figure}[htb]
\setlength{\unitlength}{1cm}
\begin{picture}(15,12)(-1,-1)
\put(-1,0){\includegraphics[scale=0.5]{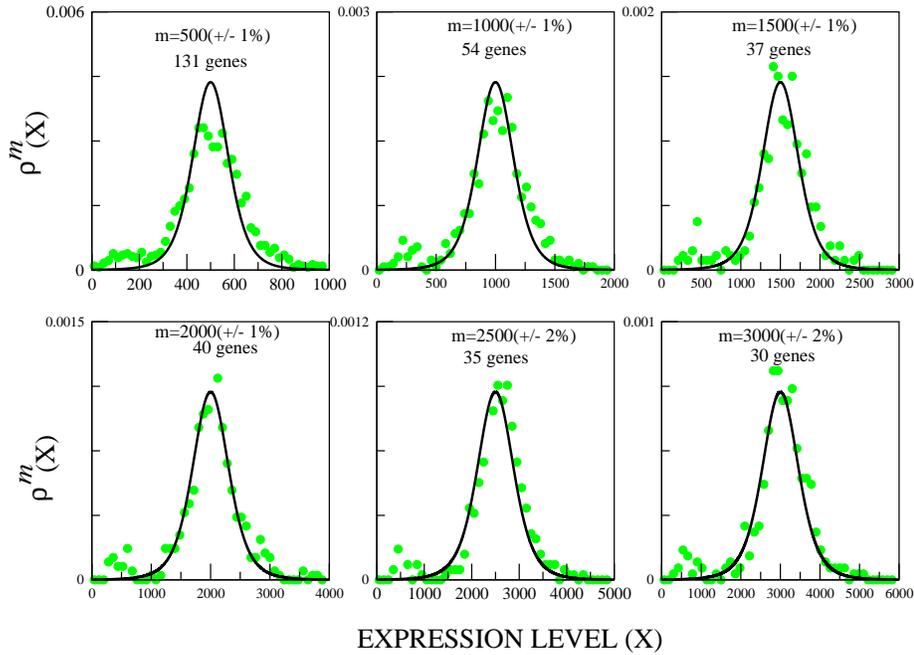}}
\end{picture} 
\caption{\small{This figure is a more detailed version of Fig. \ref{fig: probability distribution}. We show the distribution $\rho^m(x)$. It shows the distribution of expression level $x$ of genes with average expression level $m$. Horizontal axis denotes the gene expression level $x$. Vertical axis denotes the distribution $\rho^m(x)$ of genes which fluctuate around mean value $m$. 
 Green circles represent experimental data \cite{human} and solid line represents theoretical prediction in Eq. (\ref{eqn: solution of gene SPDE}). This theoretical solution (solid line) agrees with the experimental results human organism \cite{human} (green circles).
}}
\label{fig: distribution human 1}
\end{figure}

\begin{figure}[htb]
\setlength{\unitlength}{1cm}
\begin{picture}(15,12)(-1,-1)
\put(-1,0){\includegraphics[scale=0.5]{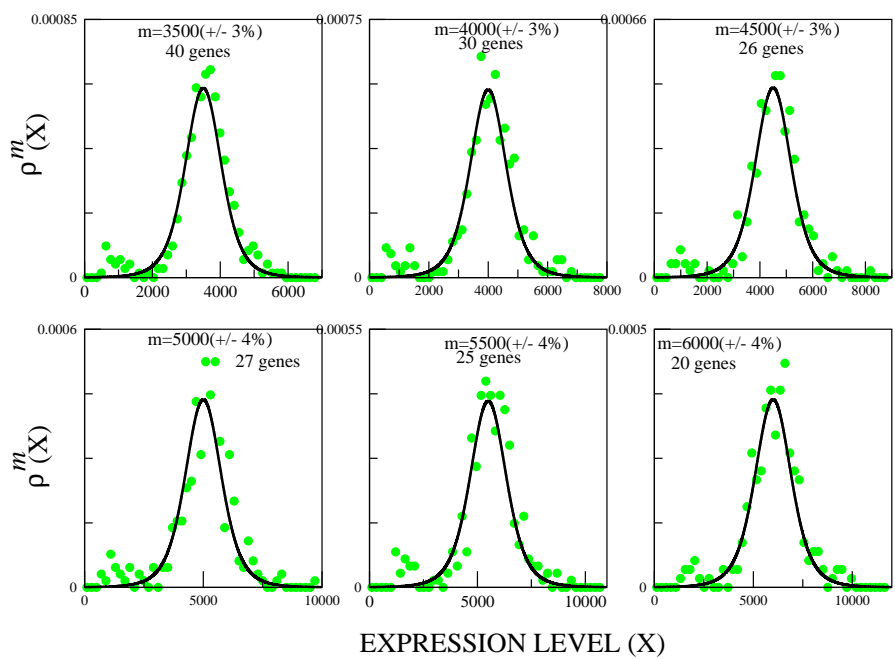}} 
\end{picture} 
\caption{\small{
Same as Fig. \ref{fig: distribution human 1} but we show genes which fluctuate around different mean value $m$.
}}
\label{fig: distribution human 2}
\end{figure}

\begin{figure}[htb]
\setlength{\unitlength}{1cm}
\begin{picture}(15,12)(-1,-1)
\put(-1,0){\includegraphics[scale=0.5]{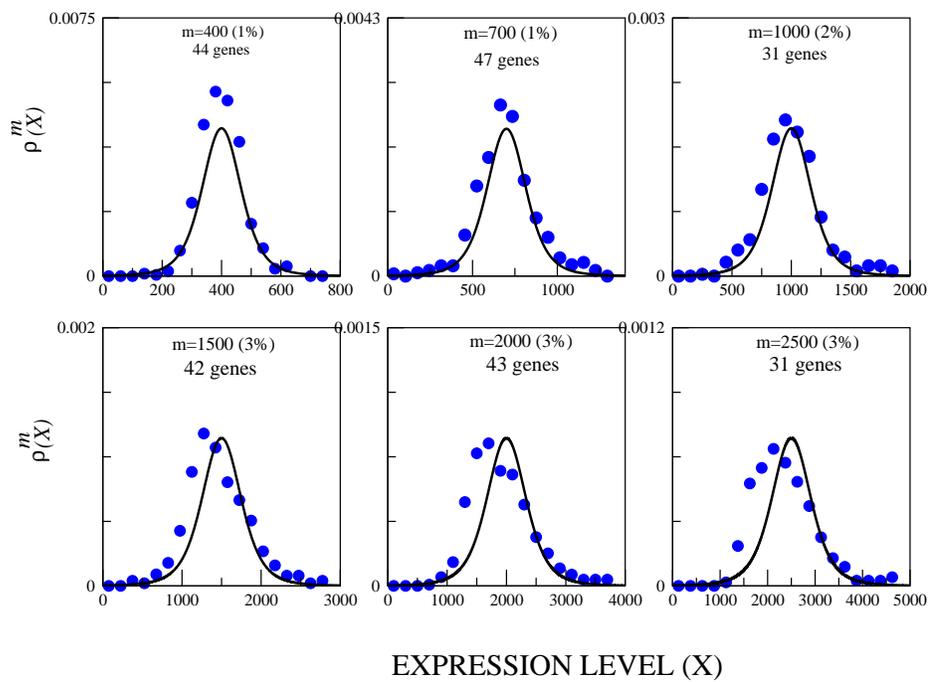}}
\end{picture} 
\caption{\small{
Same as Fig. \ref{fig: distribution human 1} but we show the results for yeast organism.
}}
\label{fig: distribution yeast}
\end{figure}

\begin{figure}[htb]
\setlength{\unitlength}{1cm}
\begin{picture}(15,12)(-1,-1)
\put(-2,0){\includegraphics[scale=0.8]{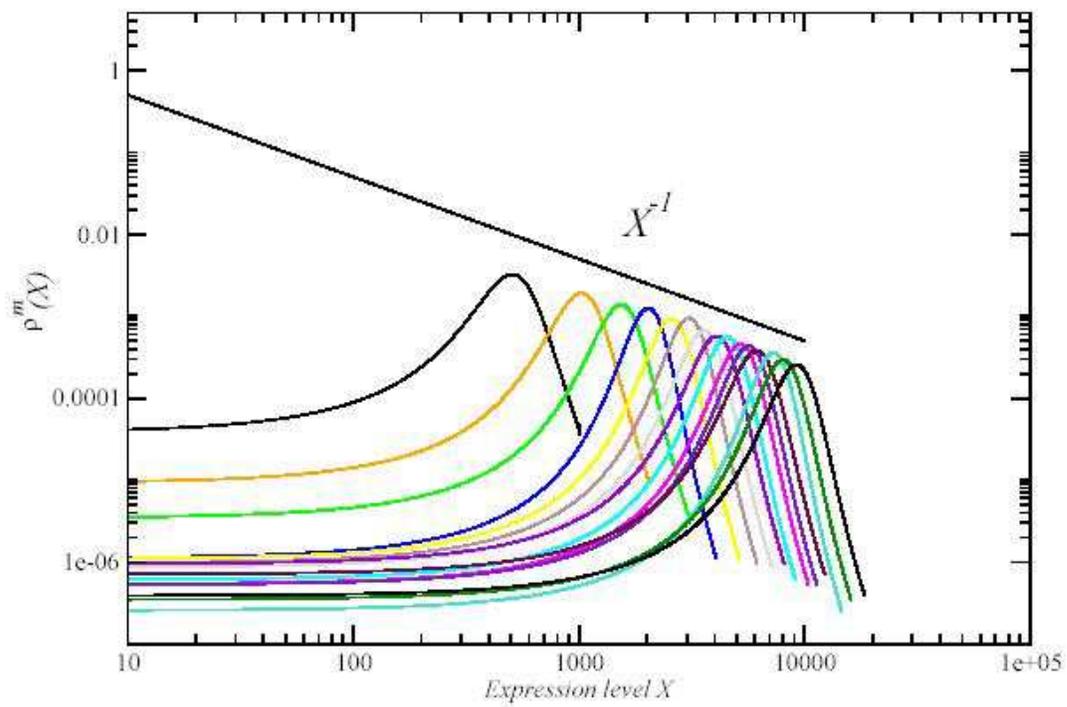}}
\end{picture} 
\caption{\small{This figure is the same as Fig. \ref{fig: probability distribution}, but in log-log scale. It is possible to observe that the convolution of the peak of distributions follows a power-law $x^{-1}$.
}}
\label{fig: peak of maximum}
\end{figure}

\paragraph{Scaling-law.}  Secondly, by direct computation, we can easily see 
that the distribution function of  Eq. (\ref{eqn: solution of gene SPDE}) has the scaling law:
\begin{eqnarray}\label{eq: scaling law}
\rho^{cm}(x)=\frac{1}{c}\rho^m(x/c)
\end{eqnarray}
(More symmetric form: $\rho^{cm}(x)dx=\rho^m(x/c)d(x/c)$) for arbitrary real number $c$.  This formula is a consequence of the self-similarity symmetry of $\alpha^m(x)$ and $\beta^m(x)$ in Eqs. (\ref{eq: symmetry drift}) and (\ref{eq: symmetry diffuion}). 
We see in Fig. \ref{fig: probability distribution} (For more  
  details see Figs. \ref{fig: distribution human 1}, \ref{fig: distribution human 2} and \ref{fig: distribution yeast}) that the experimental data follows 
this scaling-law Eq. (\ref{eq: scaling law}). The scaling law Eq. (\ref{eq: scaling law}) indicates that  the peak of each distribution is decreased by $1/m$ (see Fig. \ref{fig: peak of maximum}) and the distribution of gene expression level also  
has self-similarity structure (i.e. a repeating pattern of smaller and smaller parts 
that are like the whole, but different in size. See Figs. \ref{fig: distribution human 1}, \ref{fig: distribution human 2} and \ref{fig: distribution yeast}). 
Roughly speaking, the self-similarity structure exists in both short-time transition data (i.e., $\alpha^m(x)$ and $\beta^m(x)$) and the probability distribution (i.e., $\rho^m(x)$). They are connected with each other by means of Markov property of dynamics. Therefore, in a sense, the self-similarity governs all the dynamical structure of gene expression phenomena, by means of Markov property (i.e., all the dynamical information is scaled by the mean value $m$). 

\begin{figure}[htb]
\setlength{\unitlength}{1cm}
\begin{picture}(15,12)(-1,-1)
\put(-1,0){\includegraphics[scale=0.5]{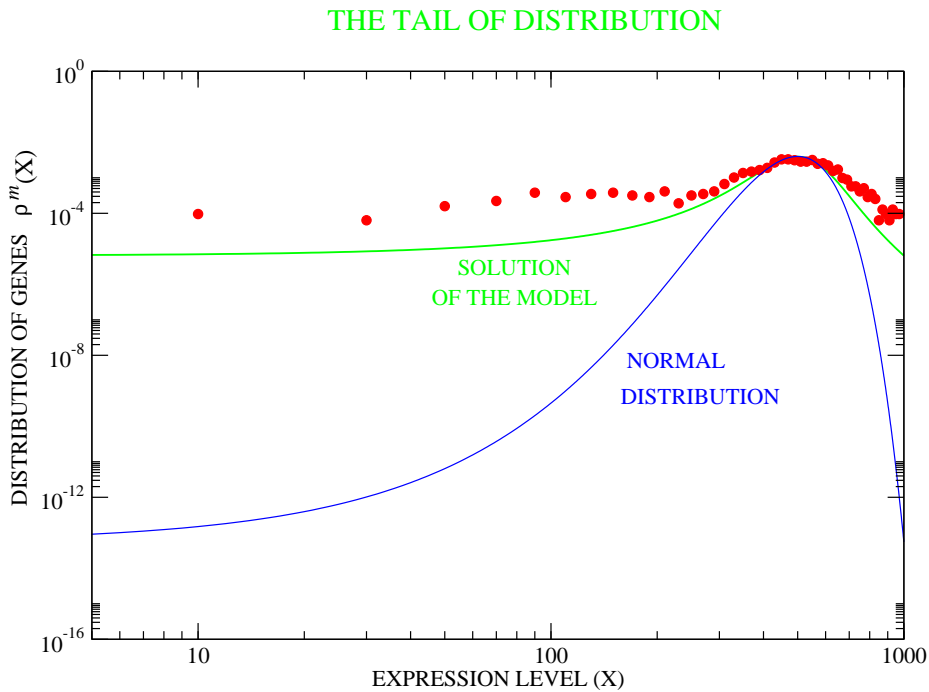}}
\end{picture} 
\caption{\small{We show the distribution $\rho^m(x)$ of expression level $x$ of genes with average expression value $m=500$ in log-log scale. Horizontal axis denotes the gene expression level $x$. Vertical axis denotes the distribution $\rho^m(x)$ of genes which fluctuate around mean value $m=500$. Dots: Experimental data of Human genes around mean value of $m=500$. Blue line: Normal distribution. Green line: The stationary distribution solution of our model Eq. (\ref{eqn: solution of gene SPDE}).  Here, we remark that the tail of distribution of experimental data obeys the power law $\rho^m(x)\propto x^{-4}$ (when $x\to \infty$), not the exponential function, as suggested by the model prediction   in Eq. (\ref{eqn: solution of gene SPDE}). This indicates that the experimental data cannot fitted by normal distribution and our model is sensitive to crucial aspects of real system.  
}}
\label{fig: Normal vs power}
\end{figure}

\paragraph{Power-law tail.} Our model can reproduce precise aspects of the gene expression system. In particular, we remark that the tail 
of the distribution solution of our model Eq. (\ref{eqn: solution of gene SPDE}) follows the power law tail $\rho^m(x)\propto x^{-4}$ (when $x\to \infty$), which is also observed in the distribution of experimental data (see Fig. \ref{fig: Normal vs power}). This indicates that the experimental distribution can not be fitted by a normal distribution, although normal 
distribution seems to be similar to the experimental distribution in linear-linear scale. The main difference is that the tail of the distribution of experimental data obeys a power-law distribution, while the tail of the normal distribution quickly decays as an exponential function. Interestingly, the experimental data indicates that the tail of distribution decays as a power-law, and not exponentially (see Fig. \ref{fig: Normal vs power}). This illustrates the powerful prediction capability of our approach because it reflects very sensitive aspects of the gene expression distribution. 

\paragraph{The peak of distribution of yeast and human.} We found that the average increasing (decreasing) tendency parameter $\mu$ in Eq. (\ref{eqn: drift}) is directly related to the maximum 
value of stationary distribution by means of an exponential function $\rho^m(m)\propto \exp(\mu/b)$ due to Eq. (\ref{eqn: solution of gene SPDE}).  By following this equation, we see that if $\mu$  increases, the maximum value of the stationary 
distribution $\rho^m(m)$ also increases.  Interestingly, this predicted   behaviour   by our model is also observed in experimental data. As we see in Figs. \ref{fig: distribution human 1}, \ref{fig: distribution human 2} and \ref{fig: distribution yeast}, the strength of the peak of 
yeast distribution ($\rho^m(m)$ for yeast) is lower than the peak of human distribution (i.e., $\rho^m(m)$ for human) because $\mu$ (yeast) $<$ $\mu$ (human).

\section{Formal analogy to Quantum Mechanics Framework}  
It is worth noticing that our approach exhibits formal analogies to the Quantum Mechanics framework. However, it is also important to remark that the following
analogies are contrained to the formal level and there is no relationship between the features of the analyzed biological system and the properties of the quantum mechanical system. But, we believe that this formal analogy will help the reader to understand more clearly the present work. 

There is one to one correspondence between our approach and the \\
Schrodinger equation and the Hamiltonian of the system as follows: 
\begin{itemize}
\item[{\it (i)}] The assumption of a stochastic process with Markov property in our 
approach can be understood as the assumption of the Schrodinger Equation in physics. 
\item[{\it (ii)}] In our approach, by using the experimental short-time transition
 probability, we characterize the dynamical equation for gene expression. This procedure 
 corresponds to identify the Hamiltonian of the system (system-dependent feature) in physics. 
\item[{\it (iii)}] Furthermore, and interestingly, the self-similarity symmetry found 
in short-time transition probability of gene expression can be understood as a symmetry 
of the Hamiltonian in physics. 
\end{itemize}

However, we remark again that despite of some formal analogies between
the Schrodinger equation and classical stochastic equations, it is also worth noticing that the difference is still crucial because quantum coherence and superposition principles are absent in our approach. 

\section{ Summary} 
\begin{figure}[htb]
\setlength{\unitlength}{1cm}
\begin{picture}(15,12)(-1,-1)
\put(-1,0){\includegraphics[scale=0.5]{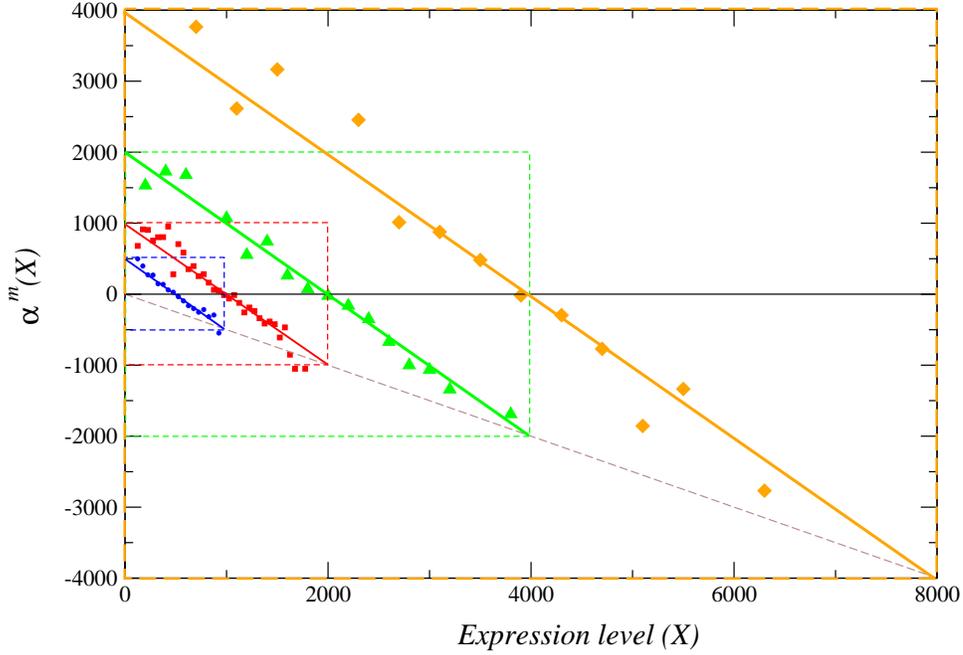}}
\end{picture} 
\caption{\small{We show the "Mean-Reverting" mechanisms. Horizontal axis denotes the gene expression level $x$.  Vertical axis denotes the drift $\alpha^m(x)$   of genes which fluctuate around mean value $m$ (i.e., genes whose average expression level is $m$).  
Orange: genes with fluctuations around expression value of $m$=4000 (resp., Green: 2000. Red: 1000. Blue: 500. ) Here, we remark that the experimental data of this figure is fitted by $\alpha^m(x)=\mu(m-x)$. This mechanism indicates that gene expression dynamics is a robust system. It means that each gene tends to recover its average value $m$. We also see that the system exhibits a self-similar symmetry (see Fig. \ref{fig: self-similarity}) because a repeating pattern of smaller and smaller parts that are like the whole, but different in size is found with respect to $m$ in this figure. (We remark that the slope $\mu$ is invariant under all $m=4000, 2000, 1000, 500$.)
This figure is constructed from selected figures with $m$=500, 1000, 2000 and 4000 in Fig \ref{fig: Mean-Reverting human}.
}}
\label{fig: Mean-Reverting}
\end{figure}

\begin{figure}[htb]
\setlength{\unitlength}{1cm}
\begin{picture}(15,12)(-1,-1)
\put(-1,0){\includegraphics[scale=0.5]{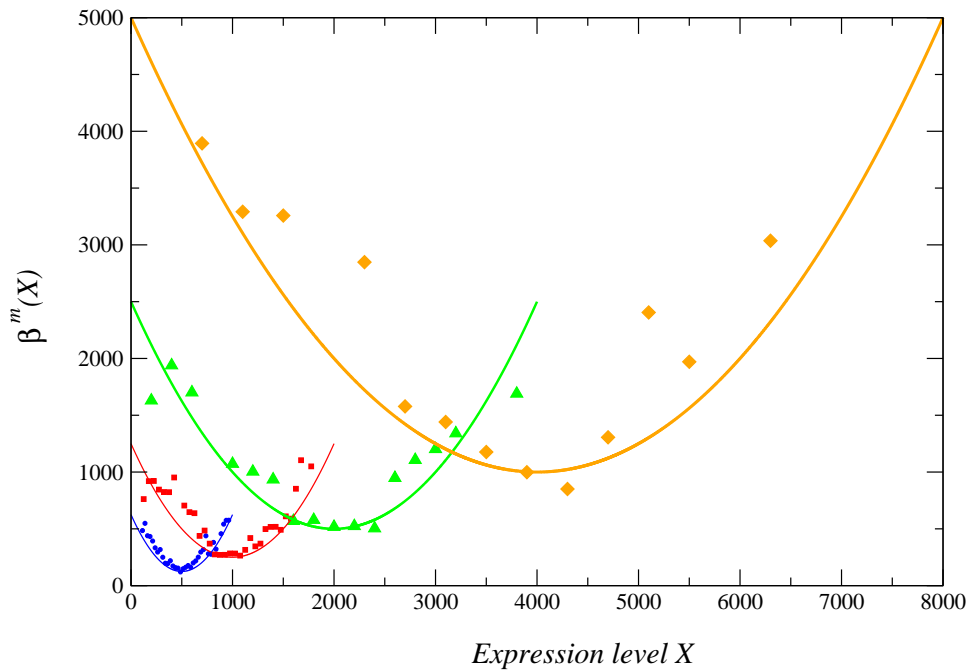}}
\end{picture} 
\caption{\small{We show "Extreme-Value-Jumps-More" mechanisms. Horizontal axis denotes the gene expression level $x$. Vertical axis denotes the diffusion $\beta^m(x)$ of genes which fluctuate around mean value $m$. 
Orange: genes with fluctuations around expression value of $m$=4000 (resp., Green: 2000. Red: 1000. Blue: 500. ) Here, we remark that the experimental data of this figure is fitted by $\beta^m(x)=m((x/m-1)^2+b)$. This mechanism indicates that values far from the mean value $m$ change more. We also see that the system exhibits a self-similar symmetry because a repeating pattern of smaller and smaller parts that are like the whole, but different in size is found with respect to $m$ in this figure. (We remark that the parameter $b$ is invariant under all $m=4000, 2000, 1000, 500$.) This figure is constructed from selected figures with $m$=500, 1000, 2000 and 4000 in Fig \ref{fig: Extreme-Value-Jumps-More human}. 
}}
\label{fig: Extreme-Value-Jumps-More}
\end{figure}

\begin{figure}[htb]
\setlength{\unitlength}{1cm}
\begin{picture}(15,12)(-1,-1)
\put(-1,0){\includegraphics[scale=0.5]{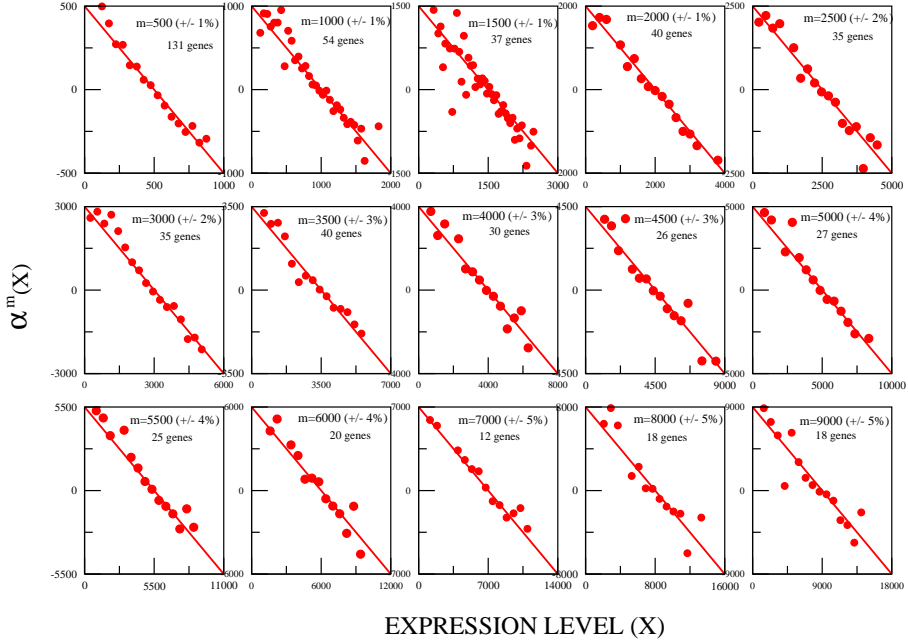}}
\end{picture} 
\caption{\small{This figure is a more detailed version of Fig. \ref{fig: Mean-Reverting}. For the drift $\alpha^m(x)$, we observe the "Mean-Reverting" mechanism in short-time fluctuation of gene expression for Human organism from $m=500$ to $m=9000$. For each figure, the vertical axis represents the drift $\alpha^m(x)$  of genes which fluctuate around mean value $m$. Horizontal axis denotes the gene expression level $x$. This mechanism indicates that genes with high expression level at time-step $t$ tend to decrease their expression level at time-step $t+1$ (i.e., $\alpha^m(x) < 0$ ), while genes with low expression level have tendency to increase their expression level (i.e., $\alpha^m(x) > 0$). This behaviour can be represented by the expression $\alpha^m(x)=\mu(m-x)$ (continuous line). Interestingly, the parameter $\mu\simeq 1$ is conserved for all the values $m$ in human organism (i.e., constant for all the subgroups of genes). Fig. \ref{fig: Mean-Reverting} is constructed from selected figures with $m=500, 1000, 2000$ and $4000$.
}}
\label{fig: Mean-Reverting human}
\end{figure}

\begin{figure}[htb]
\setlength{\unitlength}{1cm}
\begin{picture}(15,12)(-1,-1)
\put(-1,0){\includegraphics[scale=0.5]{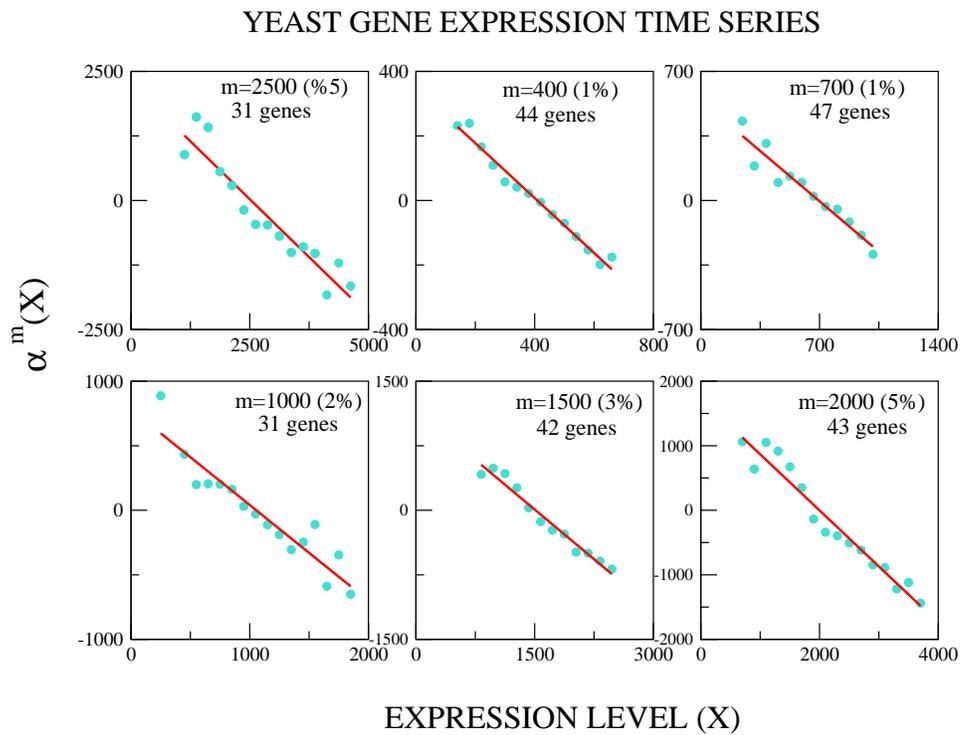}}
\end{picture} 
\caption{\small{(Same as Fig. \ref{fig: Mean-Reverting human} but this organism is Yeast) . A value in the vicinity of one $\mu\simeq 0.8$ is also observed for yeast organism . 
}}
\label{fig: Mean-Reverting yeast}
\end{figure}

\begin{figure}[htb]
\setlength{\unitlength}{1cm}
\begin{picture}(15,12)(-1,-1)
\put(-1,0){\includegraphics[scale=0.7]{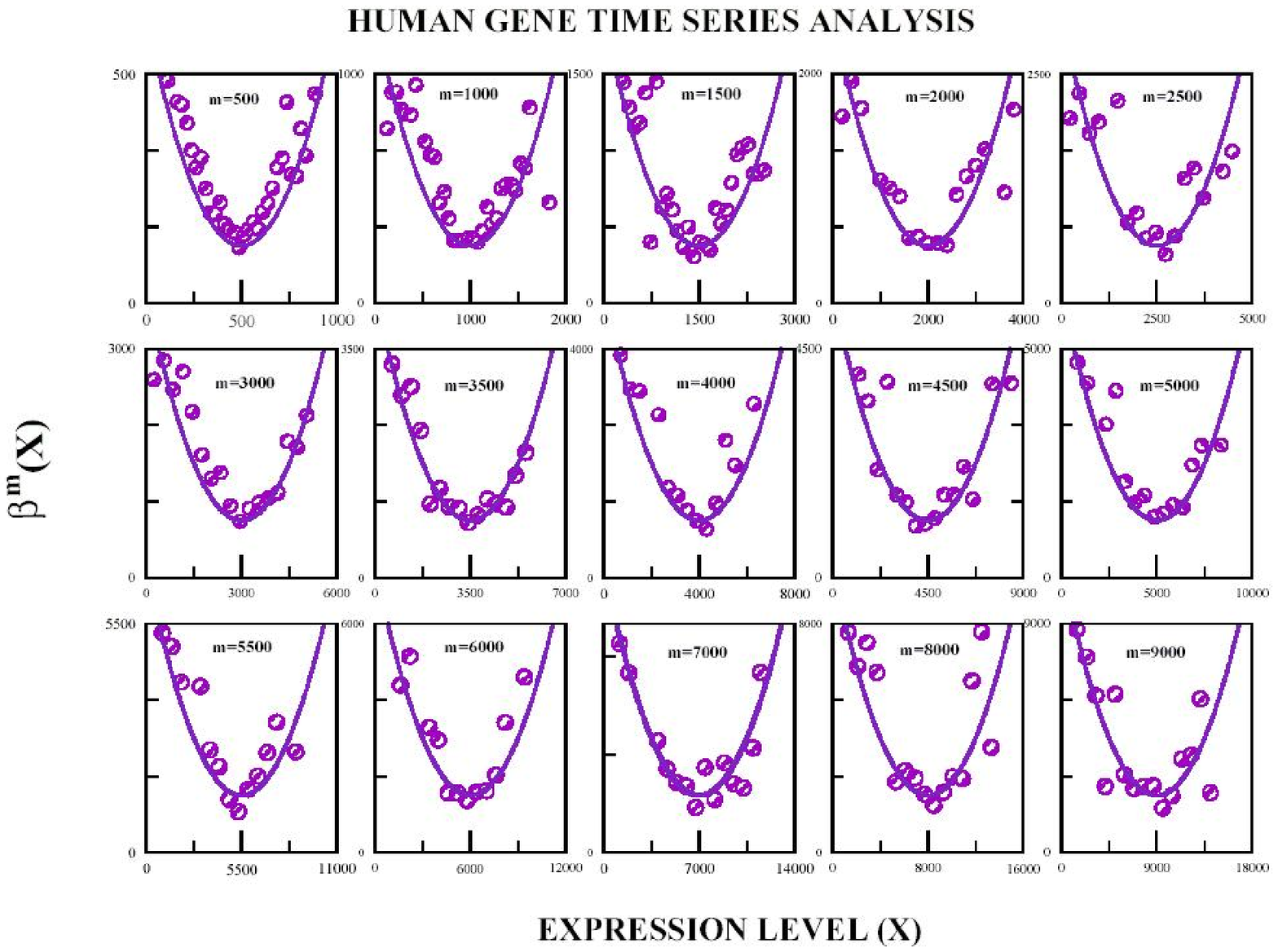}}
\end{picture} 
\caption{\small{This figure is a more detailed version of Fig. \ref{fig: Extreme-Value-Jumps-More}. For the diffusion $\beta^m(x)$, we observe the "Extreme-Value-Jumps-More" mechanism in short-time fluctuation for human organism from $m=500$ to $m=9000$. Horizontal axis denotes the gene expression level $x$ and vertical axis denotes the diffusion $\beta^m(x)$ of genes which fluctuate around mean value $m$. It means that values far from the mean value $m$ change more. This phenomena can be parameterized by the following expression $\beta^m(x)=m((x/m-1)^2+b)$. Remarkably, this value $b=0.2\sim 0.3$ is conserved under the all mean value $m$.  Fig. \ref{fig: Extreme-Value-Jumps-More} is constructed from selected figures with $m=500, 1000, 2000$ and $4000$.
}}
\label{fig: Extreme-Value-Jumps-More human}
\end{figure}

\begin{figure}[htb]
\setlength{\unitlength}{1cm}
\begin{picture}(15,12)(-1,-1)
\put(-1,0){\includegraphics[scale=0.5]{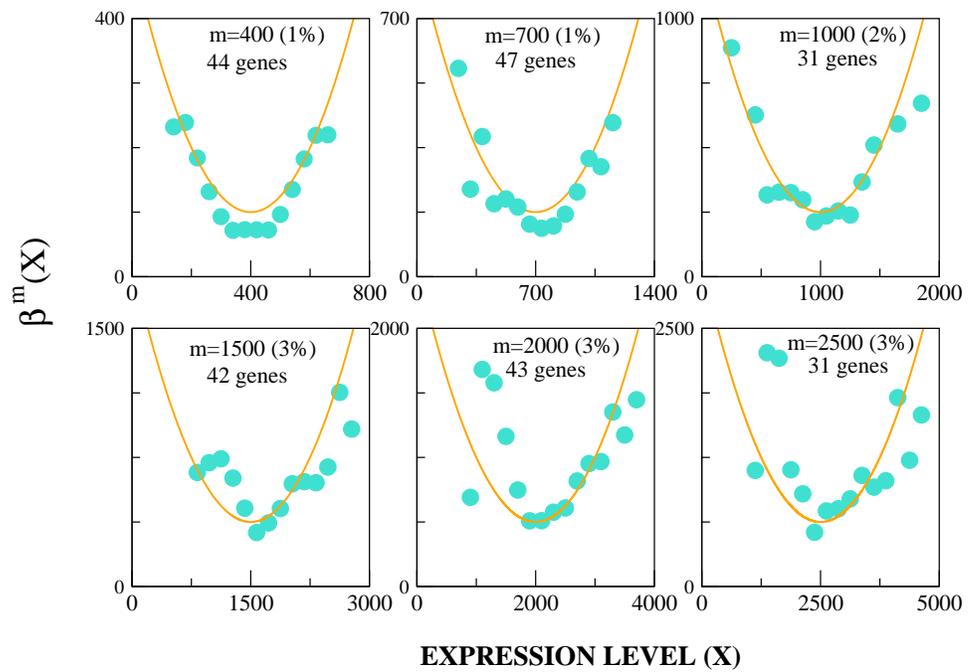}}
\end{picture} 
\caption{\small{
Same as Fig. \ref{fig: Extreme-Value-Jumps-More human} but for yeast organism.
}}
\label{fig: Extreme-Value-Jumps-More yeast}
\end{figure}

\begin{figure}[htb]
\setlength{\unitlength}{1cm}
\begin{picture}(15,12)(-1,-1)
\put(0,0){\includegraphics[scale=0.75]{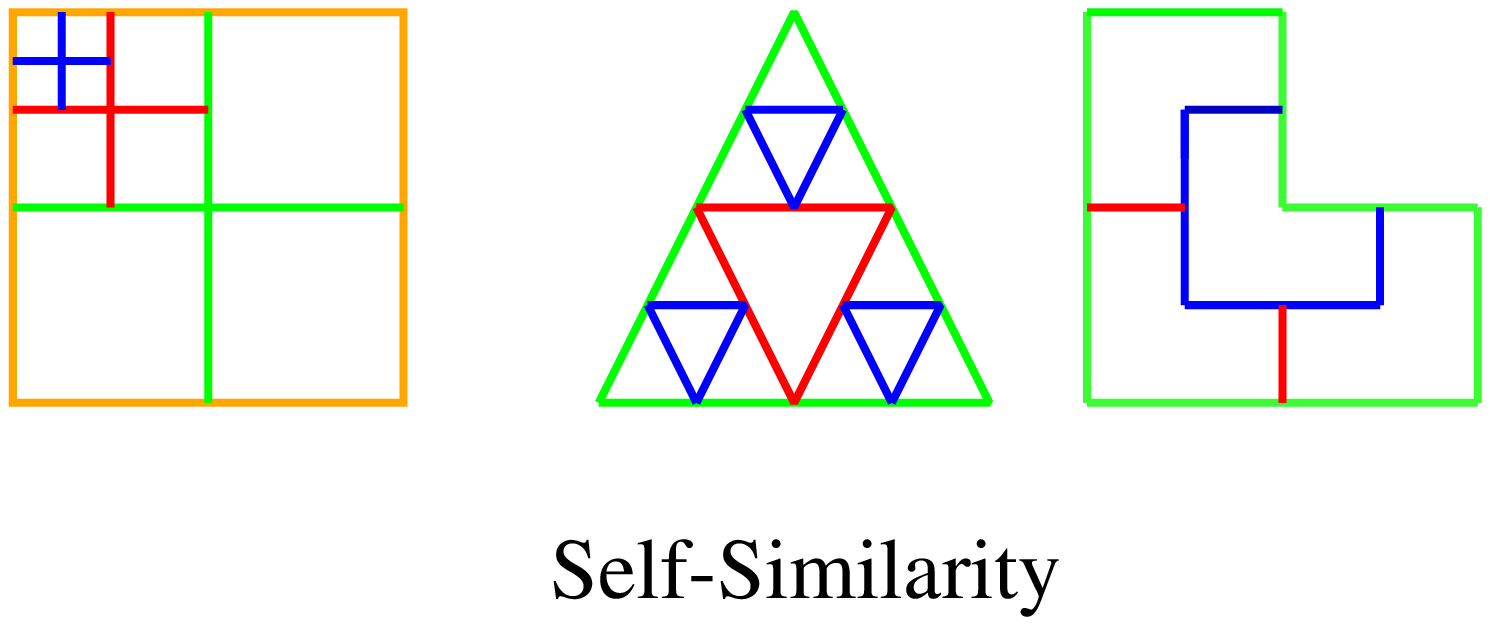}}
\end{picture} 
\caption{\small{
A figure (or system) has self-similarity if it contains a repeating pattern of smaller and smaller parts that are like the whole, but different in size. We show three examples of self-similar property.@The self-similarity property was found in gene fluctuations and it was illustrated in Figs. \ref{fig: Mean-Reverting}, \ref{fig: Extreme-Value-Jumps-More}, \ref{fig: probability distribution}, which have the same pattern to these figures. Colours have the same meaning as in Figs. \ref{fig: Mean-Reverting}, \ref{fig: Extreme-Value-Jumps-More}, \ref{fig: probability distribution}.
}}
\label{fig: self-similarity}
\end{figure}

In this article, by using the experimental data of Yeast and Human 
organisms \cite{yeast,human}, we have found the following interesting 
phenomena: 1) "Mean- Reverting" and "Extreme Value- Jumps-More" 
mechanisms. These mechanisms are present in short-time fluctuation (i.e., how the 
gene expression level changes from time-step $t$ to time-step $t+1$). Surprisingly, these two 
mechanisms are governed by a symmetry:  Self-similarity (see Figs. \ref{fig: Mean-Reverting} and Fig. \ref{fig: Extreme-Value-Jumps-More}).  By inserting this 
information (i.e., two mechanisms and the self-similarity symmetry) into the most general 
partial differential equation Eq. (\ref{eqn: general SPDE}) which spontaneously emerges from Markov property, we found the 
fundamental equation for gene expression dynamics Eq. (\ref{eqn: gene SPDE}). The stationary distribution solution Eq. (\ref{eqn: solution of gene SPDE}) for 
this fundamental equation re-builds the observed distribution $\rho^m(x)$. Finally, we succeed in naturally reconstructing the global behaviour of the observed distribution of gene 
expression (i.e., scaling-law $\rho^{cm}(x)=\frac{1}{c}\rho^m(x/c)$ ) and the precise behaviour of the power-law tail of this 
distribution $\rho^m(x)\propto x^{-4}$ (when $x\to \infty$), by using only these two ingredients: Self-similarity symmetry in short-time 
transition probability and the assumption of Markovian feature of dynamics.

The crucial point of our approach is that we can spontaneously recover both global 
behaviour of gene expression distribution (scaling-law formula) and the local behaviour 
of gene expression distribution (power-law tail), by using only two principles: Markov 
property of dynamics and self-similarity symmetry exhibited by the short-time transition 
probability.

In a sense, the self-similarity governs all the dynamical structure of gene expression
 phenomena, by means of Markov property. All the dynamical information is scaled by 
 the mean value $m$. We believe that this mechanism also happens in many other systems of Nature.

Furthermore, it is worth noticing that this self-similar symmetry may help to understand
other properties found in complex systems. For example, preliminary results shown in a companion work \cite{nacher}
indicate that the universal property of fluctuation (i.e., coupling 
between the average flux and dispersion follows a scaling-law with exponent one) reported in \cite{menezes} can be re-built
by using the self-similar symmetry explained in this work. 


Although we have solved the fundamental equation Eq. (\ref{eqn: gene SPDE}) in a stationary way, we may solve 
the same equation by searching for a time-dependent solution. This solution could predict 
the dynamical behaviour of the cells elements under some specific conditions, for example 
an external stimulus (shock) on the cell.  Theoretical studies and future experiments on 
this topic are expected to yield interesting dynamical biological knowledge.

Finally, although our analysis focuses on a single gene expression dynamics, our framework 
has a flexible capability to deal with multi-degree of freedom as it is shown in \cite{ochi2}. Similarly, the flexibility of our model allows us to extend it to other elements in 
cells as chemical compounds. Therefore, complementary time series experiments aimed to study the 
reaction fluxes and the concentrations of chemical compounds (metabolic pathways) are encouraged 
in order to obtain an integrated image of the whole cell system.
\newline

\noindent{\bf Acknowledgements}   We thank Prof.  M.J. Campbell for fruitful discussions and 
comments about experimental data. This work was partially supported by Grant-in-Aid for Scientific 
Research on Priority Areas (C) "Genome Information Science" from MEXT (JAPAN).


\newpage

\end{document}